\documentclass[preprint]{aastex631}

\usepackage{amsmath,amssymb}

\begin{document}

\title{Interstellar Dust Transport Through the Heliosphere Including the Sector Region}

\correspondingauthor{Jonathan Slavin}
\email{jslavin@cfa.harvard.edu}

\author[0000-0002-7597-6935]{Jonathan D. Slavin}
\affiliation{
Center for Astrophysics $|$ Harvard \& Smithsonian \\
60 Garden St. \\
Cambridge, MA 02138, USA
}

\author[0000-0002-3479-1766]{Marc Kornbleuth}
\affiliation{
Astronomy Department, Boston University, Boston, MA 
02215, USA
}

\author[0000-0002-8767-8273]{Merav Opher}
\affiliation{
Astronomy Department, Boston University, Boston, MA 
02215, USA
}

\author[0000-0001-8459-2100]{Gabor Toth}
\affiliation{
Department of Climate and Space Sciences and Engineering, University of Michigan, Ann Arbor, MI 48109, USA
}

\begin{abstract}
Interstellar dust has been detected \textit{in situ} flowing through the
heliosphere.  However, our ability to derive the density and size
distribution of the interstellar dust in the local interstellar medium
from this directly detected dust requires modeling the transport of the
grains as they interact with the solar wind magnetic field.  The magnetic
field in the sector region that contains the heliospheric current sheet
has rapid polarity flips which can present an effectively very low
averaged field strength to dust grains that have gyroradii tens of au in
size.  We present new calculations of dust transport through the
heliosphere using models that include the sector region to assess the
effects on dust transport. We show that the sector region can act as a
window allowing even relatively small grains to penetrate deep into the
heliosphere.  We find the sector region reduces the variation in dust
density with the solar cycle (as compared to models without the sector
region), with very little concentration or dilution of the dust for grains
larger than $\sim 0.1$ $\mu$m for most of the solar cycle. We still find a
substantial concentration of the dust in the ecliptic plane for a focusing
overall polarity of the field at solar minimum.  These models do not
include the time dependence of the magnetic field during transport of
grains through the heliosphere. Nevertheless, our results imply that
observations of interstellar dust grains, even near Earth, could be fairly
accurate in determining their size distribution in the surrounding
interstellar medium.
\end{abstract}

\keywords{Heliosphere (711) --- Interstellar dust (836) --- Solar wind (1534)}


\section{Introduction} \label{sec:intro}
Particles that penetrate the heliosphere provide us with our only direct measurements of interstellar matter. However, the complex nature of the heliosphere, with its combination of solar wind particles and magnetic field as well as various particles originating outside the heliosphere such as pickup ions, energetic neutral atoms, anomalous cosmic rays and galactic cosmic rays, makes it necessary to model the system as a whole. Interstellar dust grains are in some ways a simpler probe of the interaction of the surrounding interstellar medium (ISM) and the solar wind, since their interactions with the solar wind plasma is weak, limited primarily to charging the grains. Nevertheless, the charge on the grains results in sometimes complex interactions with the solar wind magnetic field.

On the simplest level, smaller grains (radius $\lesssim 0.05\,\mu$m  with
higher charge-to-mass ratios are excluded from the inner heliosphere. Somewhat
larger grains are substantially diverted and may be concentrated in particular
locations within the inner heliosphere, as we discuss below. The largest
grains (radius $\gtrsim 0.5\,\mu$m) suffer minimal diversion, and
their fluxes should accurately reflect the interstellar dust flux for grains
of that size. Observations that are sensitive over a range of grain sizes thus
can tell us both about the heliosphere and the ISM.

The first \textit{in situ} observations of interstellar grains in the heliosphere were made with the Ulysses spacecraft \citep{Grun_etal_1994,Frisch_etal_1999,Strub_etal_2015}. The grains identified as interstellar in origin were consistent in their speed and direction with the inflow of interstellar neutral H and the velocity vector identified for a large set of absorption lines toward nearby stars \citep{Lallement+Bertin_1992,Lallement_etal_1995,Redfield+Linsky_2008}. The main surprise in the Ulysses data was that the size distribution of the grains observed extends to large grain sizes, $\gtrsim 1\,\mu$m \citep{Landgraf_etal_2000,Kruger_etal_2015}. In addition, the size distribution of smaller grains is below what was expected based on models of dust in the general ISM. For the smallest grains, this was expected due to their small gyroradii. For the intermediate sized grains, $a \sim 0.1\,\mu$m understanding the deficit of grains requires detailed modeling. To infer the source population of grains in the ISM surrounding the heliosphere, it is necessary to understand the transport of grains into the heliosphere. It is important to use as accurate a model of the magnetic field as possible.

Although there is general agreement that the average field near the ecliptic is close to a Parker spiral, the magnetic field in the whole heliosphere volume is much more complex. One well-known feature of the solar wind magnetic field that has sometimes been left out of models for the transport of dust is the heliospheric current sheet \citep[e.g.][]{Slavin_etal_2012}. The current sheet divides regions of opposite magnetic field polarity. It is created by the combination of the overall dipole field, differential rotation of the Sun near the equator relative to the poles, and the outflow of the solar wind. Since the dominant component of the field in the wind beyond 1 au is the azimuthal component, it is the flipping of that component that can have the biggest effect on the trajectories of interstellar grains in the sector region. In addition, the tilt in the axis of the overall dipole field leads to rippling of the current sheet, the well-known ``ballerina skirt'' morphology. The rippled current sheet fills what is known as the sector region.

Shortly after the discovery of the flux in interstellar dust into the
heliosphere \citep{Grun_etal_1994}, its transport was modeled by
\citet{Landgraf_2000}. That model included the time dependence of the polarity
of the solar wind magnetic field over the course of a solar cycle, though the
field was assumed to be a Parker spiral. Grains in the simulation were assumed
to start at a distance of 100 au from the Sun and the effects of the
heliopause and termination shock were not included.
\citet{Linde+Gombosi_2000} used a 3D MHD model of the heliosphere (at
solar minimum) and started the grains outside of the heliopause but only
during the defocusing phase of the solar cycle. \citet{Czechowski+Mann_2003}
studied the effect of the rippled current sheet on grain propagation but
started the grains inside the heliopause.
\citet{Sterken_etal_2012} extended this type of modeling, carrying out Monte
Carlo simulations of dust transport to explore the effects of gravitational
force, Lorentz force, radiation pressure and grain type on grain trajectories.
That work included the time varying polarity of the solar wind
magnetic field, but not the heliosheath filtering. \citet{Slavin_etal_2012}
examined the effects of the heliospause and termination shock on grain
trajectories and resulting density distributions, using 3D
magneto-hydrodynamical heliosphere models for the magnetic field distribution
for focusing and de-focusing polarities, though with a simple planar current
sheet. Also neglected by \citet{Slavin_etal_2012} was the variation in the
solar wind magnetic field polarity over the course of a dust grain's
trajectory. This tended to  exaggerate the effect of the focusing and
de-focusing polarities on the dust distributions. With that time dependence
ignored as well as the width of the sector region, the difference in the dust
distributions for focusing and de-focusing polarities was stark with grains
even as large as 0.562 $\mu$m depleted for de-focusing polarity and grains as
small as 0.1 $\mu$m concentrated for focusing polarity. 

Recent work by \citet{Godenko+Izmodenov_2024} did include the evolving heliospheric current sheet in a realistic heliosphere model. They calculated grain trajectories in a manner similar to the way we do in this paper, though they also included the time dependence of the magnetic field during a grain's trajectory. They examined certain aspects of the resulting grain density in the region near the Sun for grains with sizes 0.15, 0.25 and 0.5 $\mu$m and found that the density of dust is enhanced in some regions near the Sun. We go into more detail in this work with regards to the effects of the sector region for grains ranging in size from 0.056 to 1 $\mu$m. In particular we illustrate how different phases of the current sheet affect the concentration of grains throughout the heliosphere, including in the outer heliosphere. While it is preferable to include the time dependence of the heliospheric magnetic field during the grain propagation, doing so would have required following a factor of several hundred more trajectories to achieve the same spatial resolution. That is because the steady flow assumption allows one to get effectively a particle for each output timestep. Without that assumption one has to follow as many particles as needed to achieve adequate statistics in each volume element of the output grid. Instead we have chosen to ignore the variation in magnetic field during propagation to see the degree to which the results are independent of time within the solar cycle.

In this paper we include the current sheet and its variations with solar cycle creating the sector region and, like \citet{Slavin_etal_2012}, start the grain trajectories far upstream such that the grains need to pass through the heliopause and termination shock to reach the inner heliosphere. We use models of the heliosphere that include the sector region in which the rippled current sheet leads to rapidly varying magnetic field directions as you go out in a radial direction. We do not include the time dependence of the magnetic field polarity and sector region during a grain trajectory, but instead focus on the possibility that the inclusion of the sector region results in much less variation with solar cycle in the dust distribution, at least close to the ecliptic plane.

\section{Methods} \label{sec:methods}
Our approach in this paper is essentially the same as in \citet{Slavin_etal_2012}. We calculate dust grain trajectories through the model heliosphere and derive the dust density in a pre-defined grid. By assuming a dust flux far upstream of the heliopause, we can then scale the derived dust density in the grid relative to the undisturbed interstellar flux. If we then assume a value for the interstellar flux based on, for example, the gas flux and the gas-to-dust mass ratio, then we can predict the absolute dust flux.

\subsection{Heliosphere Model}\label{sec:heliosphere}
We use a single-ion, multi-fluid MHD model to model the plasma-neutral interactions of the heliosphere \citep{Opher09, Opher15}. This model uses the Block-Adaptive Tree Solar wind Roe-Type Upwind Scheme (BATS-R-US) solver \citep{Toth12}, which is a 3D, block-adaptive, upwind finite volume MHD code that is highly parallel and is part of the Space Weather Modeling Framework (SWMF) \citep{Toth05}. BATS-R-US was first adapted for the outer heliosphere in \citet{Opher03}. While the global MHD model is capable of treating multiple ion species \citep{Opher20} and can be coupled to a kinetic neutral treatment \citep{Michael22, Chen24}, here we consider only one ion fluid and a four-fluid approximation for the neutrals \citep{Zank96}, as in \citet{Opher15}. A recent extension of the BATS-R-US outer heliosphere model included a level-set function to track the shape of the heliopause \citep{Onobogu24}. The level set method can be applied as a passively advected tracer, which can be attached to the plasma density to trace the interface between the outflowing solar wind plasma (assigned a value of +1 at the inner boundary) and the inflowing interstellar plasma (assigned a value of -1 at the outer boundary) \citep{Provornikova14}.

\begin{figure}[htb!]
    \centering
    \includegraphics[width=0.5\linewidth]{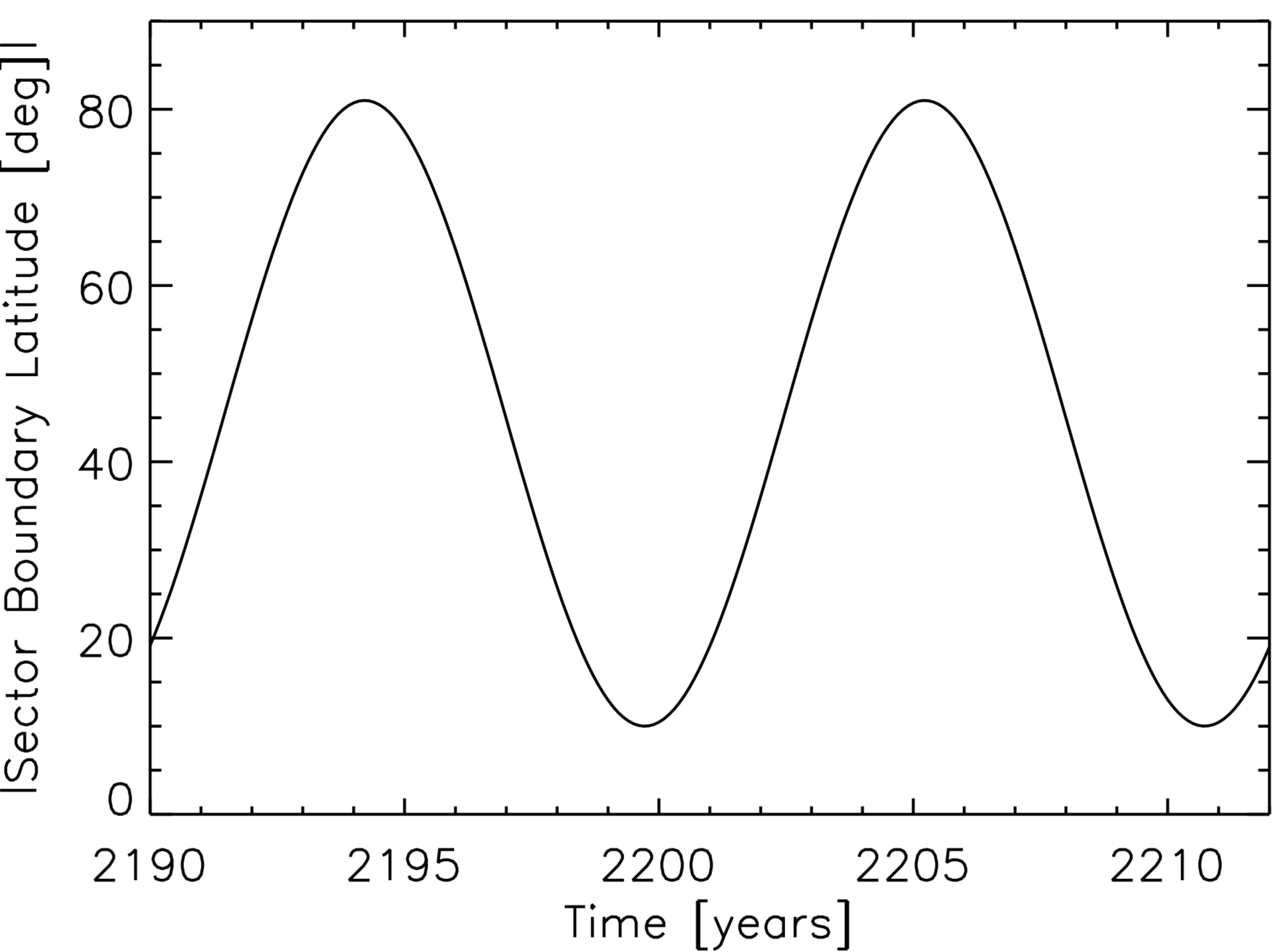}
    \caption{Time variation of the sector boundary latitude ($\alpha$) over a full 22-year solar cycle. We consider the magnitude of the latitude ($\theta$) for a given solar wind parcel at the inner boundary of 1 au with respect to the sector boundary latitude. $|\theta| < \alpha$ indicates the plasma is within the sector region, and $|\theta| > \alpha$ indicates the plasma is within a unipolar region. The time on the $x$-axis is model time and does not correspond to actual real world times.}
    \label{fig:alpha}
\end{figure}

\begin{figure}[htb!]
    \centering
    \includegraphics[width=0.65\linewidth]{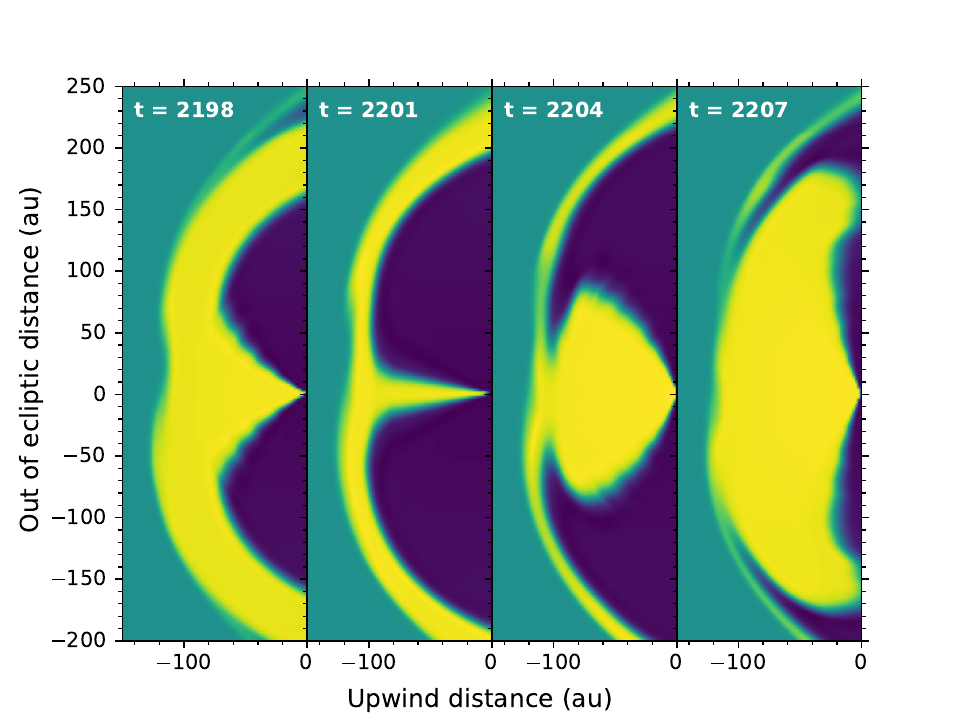}
    \caption{The sector region in a meridional plane that includes the upwind direction. Yellow regions contain the current sheet, dark blue/purple regions are inside the heliopause but do not include the current sheet and greenish regions are outside the heliopause. The times shown in the panels are model times. The solar cycle phases that correspond to the panels (l-to-r) are: transition from solar max to min, solar min, transition from solar min to max, solar max. The thickness of the sector region clearly varies strongly in time. The sector region gets advected to high and low ecliptic latitudes in the heliosheath.}
    \label{fig:curr_sheets}
\end{figure}

Here, we extend the level-set function to track the heliopause implemented in \citet{Onobogu24} to track the boundaries of the sector region caused by the folding of the heliospheric current sheet. We evaluate the evolving angle between the solar rotation and magnetic dipole axis  over the course of an 11-year solar cycle. Since we are only evaluating the sector boundaries and not the actual polarity of the solar magnetic field in our MHD model, we need only consider the 11-year cycle. We vary the sector boundary latitude ($\alpha$) over time via angle from the solar equatorial plane in our model, which is offset from the solar rotation axis by 90$^{\circ}$. The angle from the solar equatorial plane varies from 81$^{\circ}$ at solar maximum ($\alpha_{max}$) to 10$^{\circ}$ at solar minimum ($\alpha_{min}$). During an 11-year solar cycle, we therefore adopt the functional form of the variation of $\alpha$ as

\begin{equation}
\alpha = \frac{\alpha_{max}-\alpha_{min}}{2}(1 + \cos\Omega t)+\alpha_{min}
\label{eq:alpha}
\end{equation}
where $t$ is the simulation time and $\Omega=2\pi/11\,$yr. The simulation times do not correspond to actual real world times. The evolution of $\alpha$ is shown in Fig. \ref{fig:alpha}, which shows the variation of $\alpha$ between $\alpha_{min}$ and $\alpha_{max}$. We evaluate the thickness of the sector region by comparing a given latitude ($\theta$) at our inner boundary with $\alpha$. We take the magnitude of the latitude from the solar equatorial plane (${\theta}=0^{\circ}$) such that $|\theta|$ $<$ $\alpha$ indicates the plasma is within the sector region, and $|\theta| > \alpha$ indicates the plasma is within a unipolar region. The resulting current sheet evolution is illustrated in Figure \ref{fig:curr_sheets}. Note that the times for solar min and solar max include an offset of about a year to account for the time it takes for changes in the sector region to propagate out to the termination shock.

To track the sector region within our MHD, we follow a signed density function as in \citet{Provornikova14} and \citet{Onobogu24}, such that for the solar wind plasma at the inner boundary we have

\begin{equation}
F(r,t) = 
    \begin{cases}
        -\rho(r,t) & |\theta| > \alpha \\
        \rho(r,t)  & |\theta| < \alpha
    \end{cases}
\label{eq:density}
\end{equation}

\noindent and for the interstellar plasma we have $F(r,t)=-\rho(r,t)$. Here, $\rho$ is the plasma density in the MHD simulation, and the sign of the function as positive or negative indicates whether the plasma occupying a particular grid cell is within the sector region (positive) or outside of it (negative). We can evolve Eq. \ref{eq:density} throughout our domain via the passive advection equation \citep{Provornikova14, Onobogu24}

\begin{equation}
    \frac{\partial{F}}{\partial{t}}+\nabla{\cdot}(F\mathbf{v})=0
    \label{eq:continuity}
\end{equation}

By evolving $F$ with the continuity equation and following a signed density function, we are able to propagate our sector boundary parameter along plasma streamlines flowing with velocity \textbf{v} to track the evolution of the varying sector region through time and space. We can define level set function as a normalization of $F$ with respect to the plasma density considering Eq. \ref{eq:density}, such that for the solar wind plasma

\begin{equation}
f_{sector} = \frac{F}{\rho} = 
    \begin{cases}
        -1 & |\theta| > \alpha \\
        1  & |\theta| < \alpha
    \end{cases}
\label{eq:levelset}
\end{equation}

and  $f_{sector}$ = ${F}/{\rho}$ = -1 for the interstellar plasma. Therefore, for plasma tied to the sector region we have $f_{sector}$ = 1, and for plasma tied to unipolar regions we have $f_{sector}$ = -1. Due to the nature of flows within our MHD simulation, we also have grid cells containing a mixture of plasma tied to both the sector and unipolar regions. Within these cells, we are able to evaluate the dissipation of the sector region in the heliosheath. The mixing can be evaluated as

\begin{equation}
    f_{sector} = \frac{F_{sec}+F_{uni}}{\rho_{sec}+\rho_{uni}} = \frac{F_{sec}+F_{uni}}{\rho} = \frac{\rho_{sec}-\rho_{uni}}{\rho}
\end{equation}

\noindent where $F_{sec}$ and $F_{uni}$ are the signed density functions of the plasma tied to sector and unipolar regions, respectively, and $\rho_{sec}$ and $\rho_{uni}$ are the densities of the plasma tied to sector and unipolar regions, respectively. These plasma parcels can be combined within a grid cell as multiple streamlines converge within a single cell. Therefore, when $\rho_{sec} > 0$ and $\rho_{uni} > 0$ within a grid cell, then $-1 < f_{sector} < 1$.

For the outer boundary of the MHD model, all of the neutral and ionized populations in the interstellar medium are assumed to have the same bulk velocity $v_\mathrm{ISM}$ = 26.4 km/s (longitude = 75.4$^{\circ}$, latitude = -5.2$^{\circ}$ in ecliptic J2000 coordinate system) and temperature $T_\mathrm{ISM}$ = 6519 K at the outer boundary 1500 au from the Sun, where the pristine ISM is not mediated by the heliosphere. In the ISM, the proton density is assumed to be $n_\mathrm{p,ISM}$ = 0.06 cm$^{-3}$ and the neutral H atom density is $n_{H,ISM}$ = 0.18 cm$^{-3}$, which is in agreement with the pristine neutral H atom density inferred by \citet{Swaczyna20} based on New Horizons observations ($n_{H,ISM}$ = 0.195 $\pm$ 0.033 cm$^{-3}$). For the interstellar magnetic field ($B_{ISM}$), we use $B_{ISM}$ = 3.0 $\mu$G, with $\alpha_{BV}$=15$^{\circ}$, where $\alpha_{BV}$ is defined as the angle ($\alpha_{BV}$) between B$_{ISM}$ and the interstellar flow in the hydrogen deflection plane \citep{Lallement05, Lallement10}. This orientation and magnitude of $B_{ISM}$ is derived from the $B_{ISM}$ trends noted to best match interstellar Ly-$\alpha$ absorption observations\citep{Powell24} and observations of energetic neutral atom profiles of the heliotail \citep{Kornbleuth24}.

For the inner boundary conditions, the MHD model uses 22-year averaged solar cycle conditions (1995-2017) from \citet{Izmodenov20} and \citet{Kornbleuth21}. The model takes into account helio-latitudinal variations of the solar wind density and speed \citep{McComas00,Sokol13,Tokumaru21}, and the temperature is related to the solar wind speed via an assumed sonic Mach number of $M=6.44$ at Earth (corresponding to a solar wind temperature of $T_{SW}$ = 188,500 K). In the ecliptic plane, the solar wind speed and density are derived from hourly-averaged solar wind data from the OMNI 2 dataset \citep{King05}. Heliolatitudinal variations of the solar wind speed and density are also considered based on interplanetary scintillation (IPS) observations \citep{Tokumaru12} and variations of the solar wind mass flux from SOHO/SWAN full-sky maps of backscattered Lyman-alpha intensities are used \citep{Quemerais06, Lallement10,Katushkina13, Katushkina19}. 
The solar wind mass flux as a function of time and heliolatitude is obtained via an inversion procedure. The solar magnetic field is derived from a Parker solution with the strength of the radial component being $B_{SW}$ = 37.5 $\mu$G at 1 au.
Therefore, while a steady state solution is obtained for the MHD plasma solution, the time-varying sector boundary is superposed on top of the static solution. A more accurate method for modeling the time-varying sector boundary would be in conjunction with a time-dependent model of the heliosphere. However, for the qualitative purposes of this work the more accurate form of the time-dependent modeling of the solar wind plasma is not required.

The computational domain for the MHD model is $x,y,z=\pm1500$ au. The coordinate system is such that the z-axis is parallel to the solar rotation axis and the x-axis is aligned with the longitude of the inflow but is 5$^{\circ}$ above it, with y completing the right-handed coordinate system. Our inner boundary is set as a sphere at 1 au, with resolution of 0.0625 au. We use a resolution of 1 au or less from $x=-16$ to 16 au, $y=-16$ to 16 au, $z=-16$ to 16 au. In the supersonic solar wind and heliosheath near the termination shock we use a resolution of 2 au from $x=-190$ to 400 au, $y=-200$ to 200 au, $z=-200$ to 200 au. In the heliosheath and extending into the ISM we use a 4 au resolution (from $x=-306$ to 1006 au, $y=-476$ to 476 au, $z=-476$ to 476 au). 

\begin{figure}
    \centering
    \includegraphics[width=0.4\linewidth]{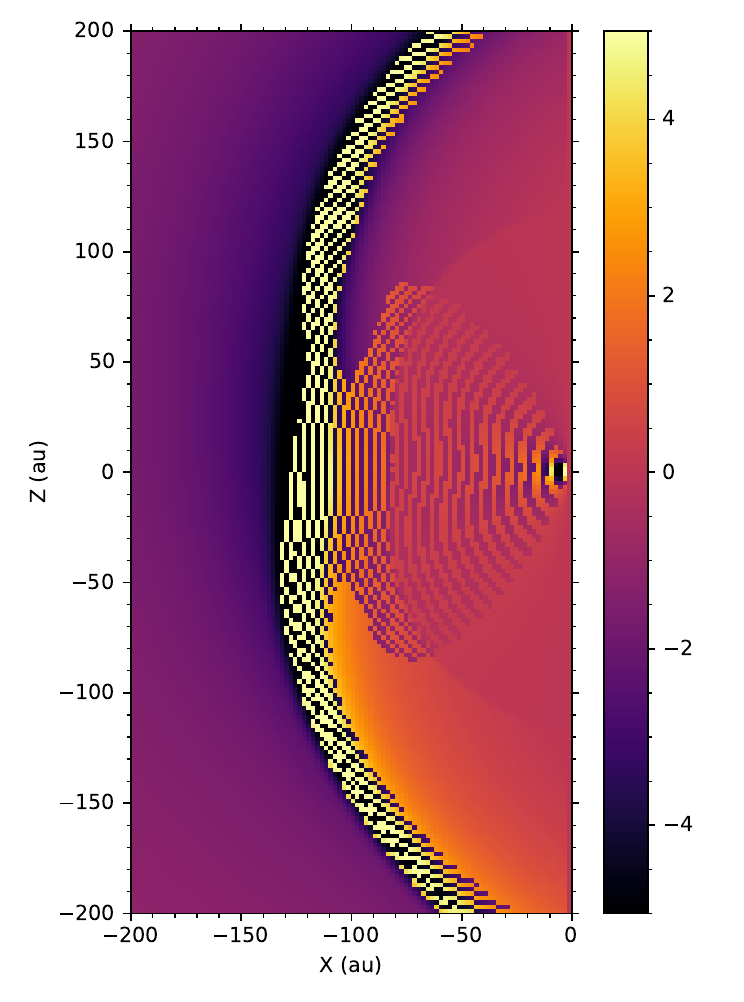}
    \caption{Magnetic field, $B_y$ (in $\mu$G), for the time of transitioning from solar min to solar max (model time 2204) in the meridional plane that includes the upwind direction. The overall polarity imposed is focusing (magnetic north in the ecliptic north). The azimuthal field in this plot is $-B_y$.}
    \label{fig:currsheet_Bfield}
\end{figure}

For the purposes of modeling the dust transport, the polarity of the magnetic field is important. Therefore we have altered the unipolar field in the models to create a bipolar overall field with either focusing or defocusing polarity. We do this by reversing the magnetic field direction, i.e.\ flipping the sign of all of the components of the field, inside of the heliopause in either the ecliptic north or the ecliptic south hemispheres. To simulate the rapidly fluctuating field in the sector region, in addition to the changes for overall polarity, we impose reversals of the field in the sector region that depend on distance from the Sun as $f(r) = \sin(2\pi r/R_{rot})$
where $R_{rot} = v_r \tau$ is the approximate distance that the wind flows
radially during one rotation period, $v_r$ is the radial speed of the solar
wind, $\tau$ is the Sun's rotation period and $r$ is the radial distance from
the Sun. Where $f(r) > 0$ we reverse the magnetic field and otherwise leave it
as it was. There is no dependence on latitude assumed. This is clearly an
approximate treatment, but is adequate for our purposes. We note that the
spatial period of these reversals is close to the spatial resolution of the
heliosphere model that we use in many regions, thus a more gradual flipping of
the field is not feasible. We limit the period of the reversals to be no less
than the spatial resolution of the model grid, which prevents aliasing. Since
the grains cross a single spatial zone in a very short time, and in general
have gyroradii at least several times larger than the spatial variation of the
field, they experience very little deflection over a single grid element.
Therefore we do not see this limitation of the scale of the field reversals as
a significant limitation. The resulting field strength in the azimuthal
direction is shown for one particular case in Figure
\ref{fig:currsheet_Bfield}. The case shown is for model time 2204, which
corresponds to a time of transition from solar min to solar max. The overall
polarity is focusing.

We note that the distance to the termination shock and, to a lesser degree,
the heliopause vary over the course of the solar cycle. This variation could
influence dust trajectories but is not taken into consideration during the
dust transport calculations described below.

\subsection{Dust Model}\label{sec:dust}
The critical features of the dust model for our calculations are those that
determine the charge-to-mass ratio for the grains. These include various
parameters that determine the charge on the grains and the mass density of the
grain material. In this paper we focus on silicate grains. It has long been a
standard assumption that grains are either silicates or carbonaceous. This
assumption has recently been questioned \citep{Hensley+Draine_2023}, however
both absorption line data \citep{Slavin+Frisch_2006ApJ} and dust detections by
the Cassini spacecraft \citep{Altobelli_etal_2016} tend to favor little to no
carbonaceous dust in the very local ISM. For the grains in our model we assume
a density of the solid grains of 3.3 g cm$^{-3}$. This is a fairly typical
value to assume \citep[for example][]{Jones_etal_1994}, though lower values,
e.g.\ 2.65 g cm$^{-3}$, have been found depending on the assumed composition.
Some interstellar grains captured by the Stardust mission
\citep{Westphal_etal_2014} had low bulk densities,
\citet{Alexashov_etal_2016} assumed a density of 2.5 g cm$^{-3}$.

To calculate the charge on the grains we need to provide the UV radiation
field. For that we reuse the one from \citet{Slavin_etal_2012}
which is made up of the constant FUV interstellar radiation field from \citet{Gondhalekar_etal_1980} and the solar radiation field, which goes as $1/r^2$. 
The EUV/FUV part of the solar radiation field is taken from the observed and modeled data provided by
the Solar EUV Experiment (SEE) on the NASA TIMED (Thermosphere Ionosphere Mesosphere Energetics and Dynamics,
\url{http://lasp.colorado.edu/see/}) mission, which provides daily EUV/FUV spectra based on solar observations and models. For the visible/near UV we use data from the SORCE (Solar Radiation \& Climate Experiment,
\url{http://lasp.colorado.edu/sorce/}) mission.

The grain charging calculations use optical constants from \citet{Weingartner_etal_2006}, calculated with code generously provided to us. We assume that the charging is in equilibrium, i.e.\ that the charge is that which comes from assuming that the positive charging rates, by sticking of protons, photoionization and secondary electron ejection, balance with the negative charging rates by electron sticking. For small grains, \citep[$\lesssim 0.01\,mu$m][]{Godenko+Izmodenov_2023}, this assumption can be violated, however most such grains are excluded from the inner heliosphere in any case.

\subsection{Dust Trajectory Calculations}\label{sec:traj}
We start the dust trajectory calculations with a grid of grains at a location 600 au upstream of the Sun. The coordinate system of the heliosphere model is oriented with $-x$ as the direction in the ecliptic plane at the inflow ecliptic longitude, $z$ is toward ecliptic north and $y$ completes the right handed system. Thus the interstellar flow is nearly in the $+x$ direction, though with a small $-z$ component, since the flow comes from a direction about 5$^\circ$ above the ecliptic plane. The initial positions of the dust grains are in a plane at $x = -600$ au in a grid that has finer resolution than the output grid, 0.3 au in each direction. With a total $1001^2$ particles, the initial grid is 300 au on a side. The grid is centered such that the grain trajectories fully cover the output grid. In other words, we make sure that the output densities are not affected by lacking input grain positions. The 3-D output grid is 200 au on a size with voxels that are 4 au on a side, going from $x = -200$ au to 0 and from $-100$ to $+100$ au in $y$ and $z$.

The grains start with a speed equal to that of the gas at their location, though with an additional component perpendicular to the magnetic field at that location. The magnitude of the additional component is 1 km $s^{-1}$ and it is oriented randomly in the plane perpendicular to the field. Thus the grains start with a small gyrovelocity. The trajectories of the particles are then calculated by integrating the equations of motion including the Lorentz force, gravitational and radiation pressure force. The integrations are done with a constant output time step using the DVODE code \citep[][available from \url{https://computing.llnl.gov/projects/odepack/software}]{Brown_etal_1989}. The time step is chosen so as to get several hits per voxel as a grain traverses one, $dt = 1.73\times10^6$ s, though this value doesn't affect the results unless it is so large that voxels get a small number of hits. After each output time step the dust number value for the voxel containing the dust particle is incremented. After all trajectories have been calculated, comparison of the values in the output array with those in the most upstream voxels, i.e. at $x = -600$ au, we can get the density enhancement factor. The number of hits in a voxel with ambient density should be equal to the value $1/(dy_g\, dz_g\, v_{x0}\, dt)$, where $dy_g$, $dz_g$ are the grid sizes in the $y$ and $z$ directions and $v_{x0}$ is the initial grain velocity in the $x$ direction, though noisiness can cause small differences from that value. 

\begin{figure}[htb!]
    \centering
    \includegraphics[width=0.5\linewidth]{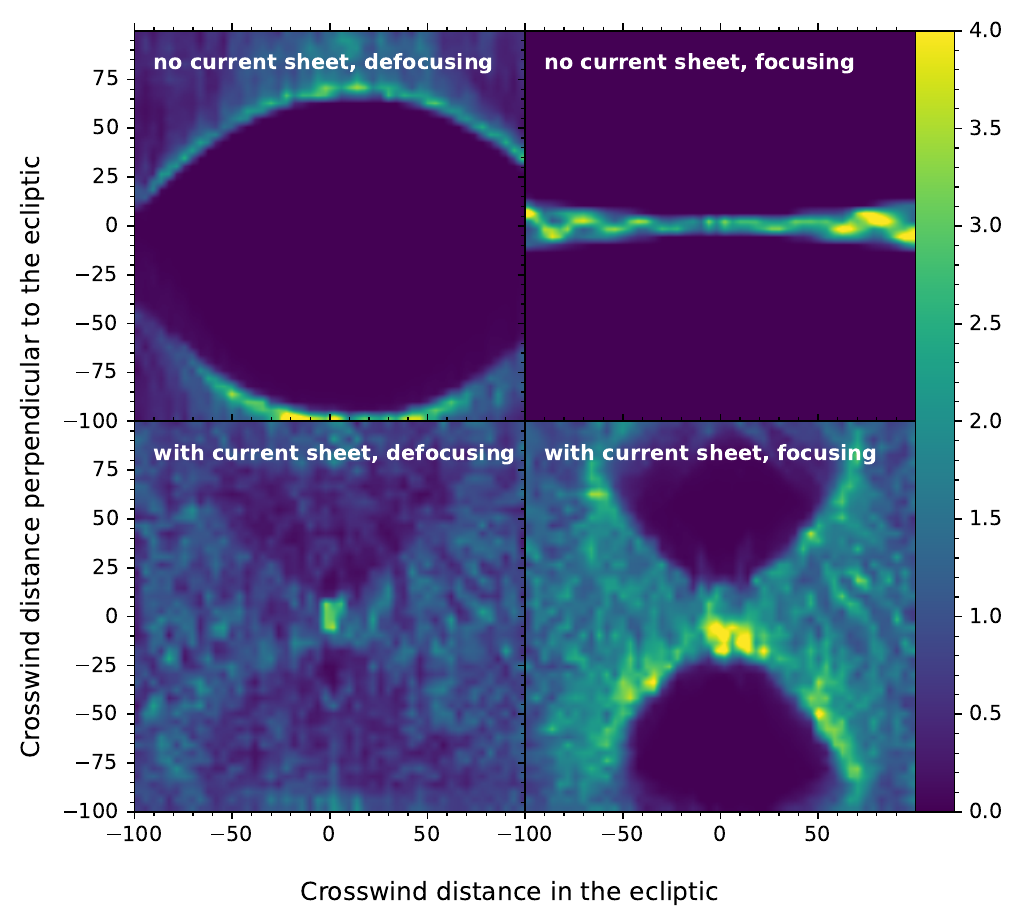}
    \caption{Comparison of dust density relative to interstellar for 0.1 $\mu$m grains and for different assumptions about the magnetic field. The images are for a slice perpendicular to the ecliptic at a distance 20 au upstream of the Sun. As indicated by the labels, the heliosphere model for the calculation either included the sector region or did not and the overall polarity is either focusing or defocusing. Clearly there is much less of a difference between the focusing/defocusing cases when the sector region is included.}
    \label{fig:comp_wcurrsheet}
\end{figure}

We have done the trajectory calculations for the four different model times as shown in Figure \ref{fig:curr_sheets} and with either focusing or defocusing overall polarity to the field. Thus there are eight calculations per grain size. We calculated trajectories for six different grain sizes over a logarithmically spaced grid ranging from $\log_{10}(a(\mu\mathrm{m})) = -1.25$ to 0, i.e. from $a= 0.056$ to $1\,\mu$m.
This range was chosen because smaller grains would be completely excluded from the inner heliosphere and larger grains would be essentially undisturbed from their initial trajectories. With the six different grain sizes and effectively eight different heliospheric magnetic field configurations, we calculated a total of 48 models for the dust trajectory/density distributions.

\begin{figure}[htb!]
    \centering
    \includegraphics[width=0.75\linewidth]{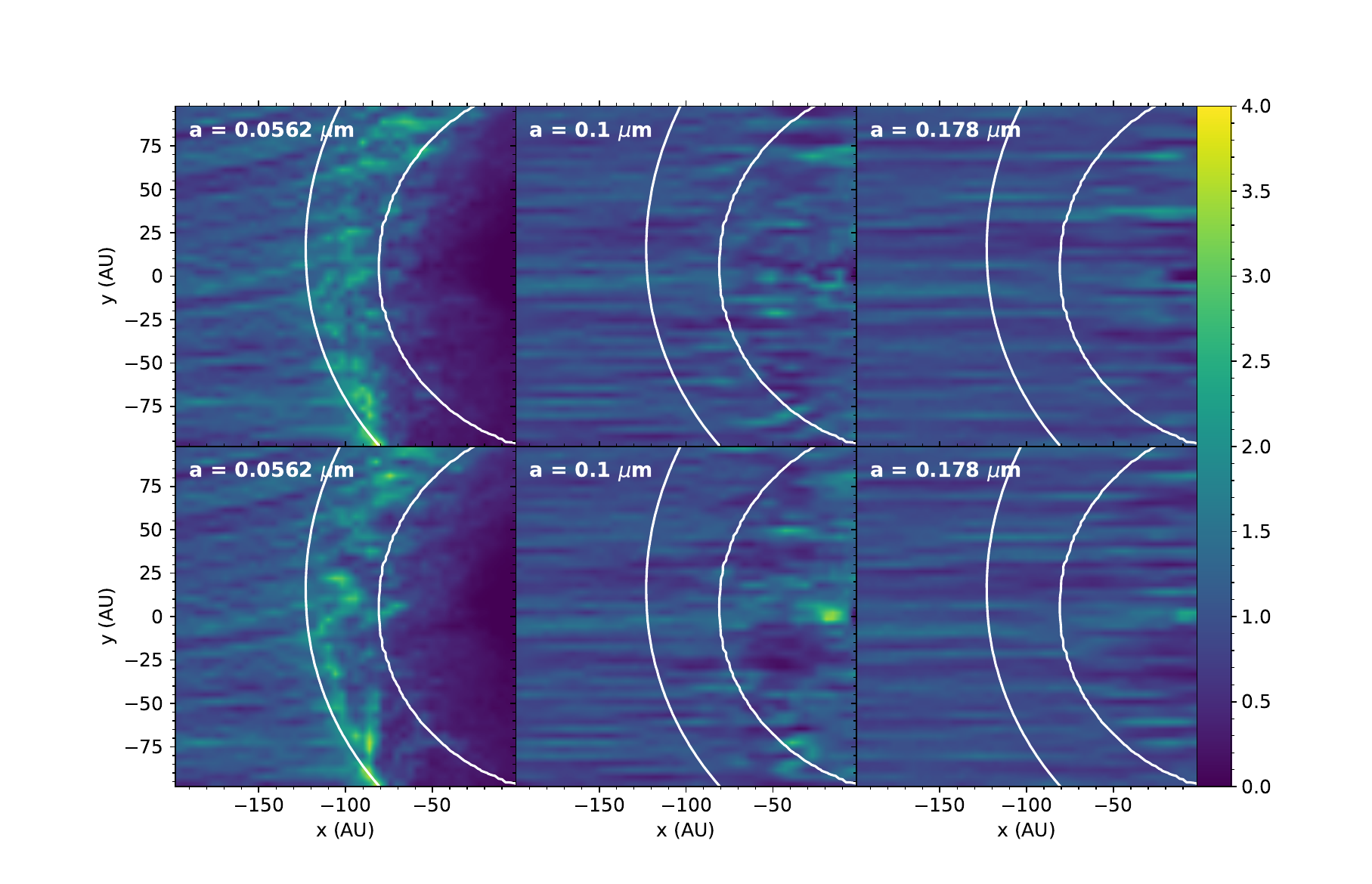}
    \caption{Grain density relative to that in the undisturbed ISM in the ecliptic plane. The field here is for solar max conditions (model time 2207). The top row is for overall focusing polarity and the bottom row is for defocusing polarity. The grain sizes are as indicated in the labels. The white lines indicate the location of the termination shock (to the right) and heliopause (to the left).}
    \label{fig:ddens_small_solmax}
\end{figure}

\begin{figure}[htb!]
    \centering
    \includegraphics[width=0.75\linewidth]{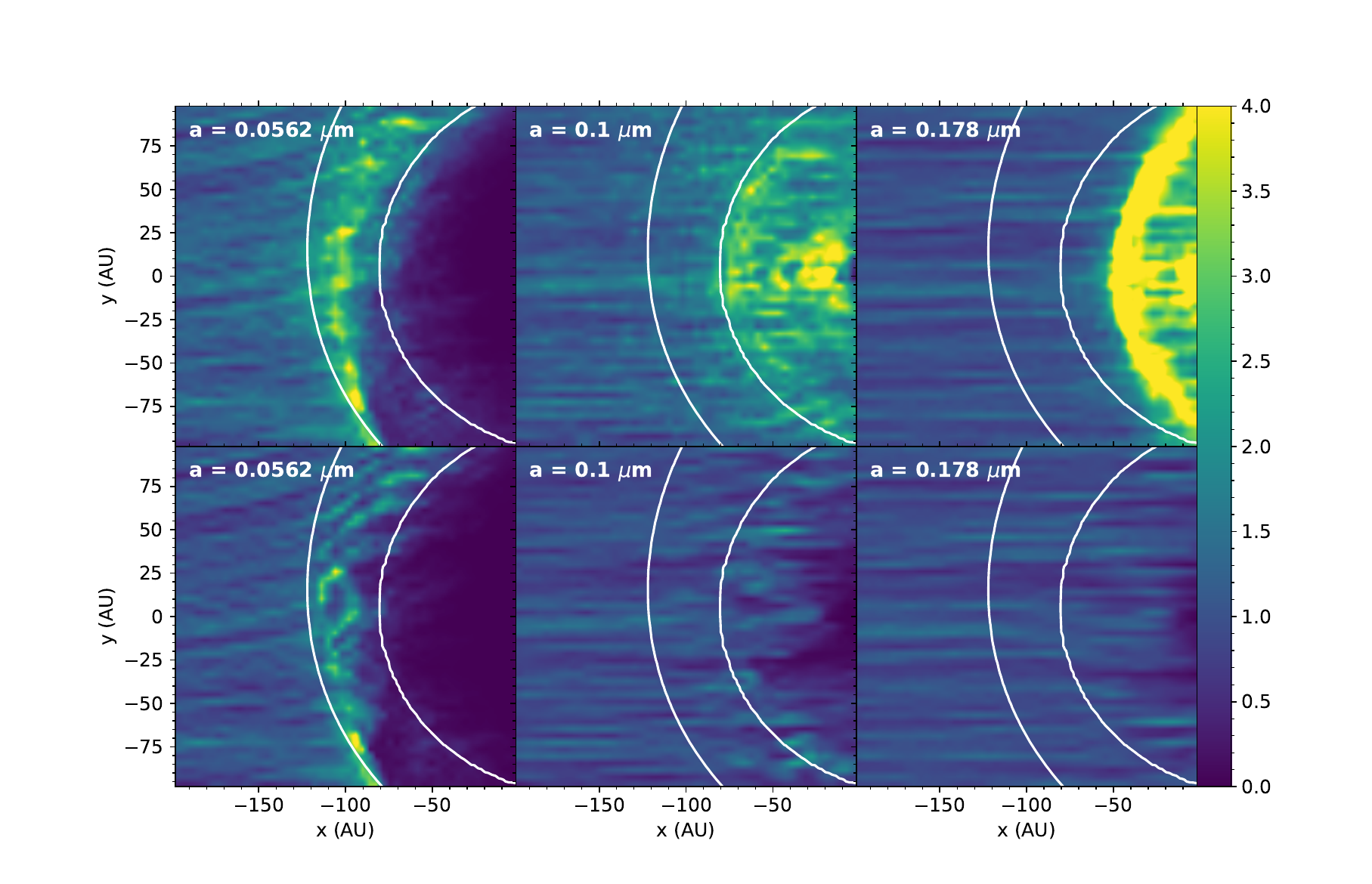}
    \caption{Same as Figure \ref{fig:ddens_small_solmax} except that the sector region is that for solar minimum (model time 2201, see Figure \ref{fig:curr_sheets}). The dust density is clearly much more disturbed from ambient density than for solar max conditions, especially for the focusing polarity (top row).}
    \label{fig:ddens_small_solmin}
\end{figure}

\section{Results}
To assess the importance of the presence of the sector region with its rippled
magnetic field we calculated grain trajectories with the model as described
with an overall polarity, either focusing, i.e. magnetic north aligning with
ecliptic north, or de-focusing, the opposite polarity, and compared with the
same model with the addition of the rippled magnetic field in the sector
region. The results for the dust density distribution for 0.1 $\mu$m grains
are dramatic as shown in Figure \ref{fig:comp_wcurrsheet}. The differences of
the focusing vs. de-focusing polarities are much less significant when the
sector region is included, though differences do remain. Note that the slice
in this figure is at 20 au upstream of the Sun. The blobs in the lower panels
(which include the sector region) are the result of focusing by the heliopause
and termination shock and are parts of filaments that do not reach to the Sun.
Slices in the ecliptic can be seen in Figure \ref{fig:ddens_small_solmax}
center panels (top and bottom).

In Figures \ref{fig:ddens_small_solmax} and \ref{fig:ddens_small_solmin} we
show the dust density relative to ambient for solar max conditions and solar
min conditions respectively. In these figures the density is for a slice in
the ecliptic plane. (Note that the streaks are an artifact that arises
because, with the steady flow assumption, ``particles'' correspond to points
equally spaced in time along a trajectory and thus are smooth in the direction
of the trajectories, while  numerical noise exists between trajectories.) The
termination shock and heliopause are indicated by the white lines. It
can be seen that the smallest grains are somewhat deflected upstream of the
heliopause. This is because of the interstellar magnetic field that wraps
around the heliosphere and is able to deflect small grains within the region
in which that distortion of the field begins. The two cases shown in Figure
\ref{fig:ddens_small_solmin} show the most contrast of any that we calculated.
Clearly solar min conditions, with its thin sector region, is most like
previous calculations that used a planar current sheet. Comparing with
\citet{Slavin_etal_2012}, their figure 7 compares with our figure
\ref{fig:ddens_small_solmin}. The de-focusing polarity cases do not show as
much depletion for our calculations since even for solar minimum conditions
there is a thin current sheet that leads to a low effective magnetic field.
For the focusing polarity we see a large density enhancement upstream as in
\citep{Slavin_etal_2012} because grains away from the ecliptic are outside of
the current sheet and get focused toward the ecliptic. We note that these
large concentrations of dust exaggerate the effects of the magnetic field
because we do not include the time dependence of the field during a grain
trajectory. For a grain moving at 26 km s$^{-1}$ it takes roughly 22 years to
go from the heliopause (assuming a distance of 120 au) to the Earth. During
that time an entire solar cycle will occur. However, as we show below, only
for the period near solar minimum does the field substantially alter the dust
density. The variation with time of the polarity will lead to
alternating periods of focusing and de-focusing, which will tend to smooth out
the variations in dust density such as in the upper right panel of Figure
\ref{fig:ddens_small_solmin}.

\begin{figure}[htb!]
    \centering
    \includegraphics[width=0.65\linewidth]{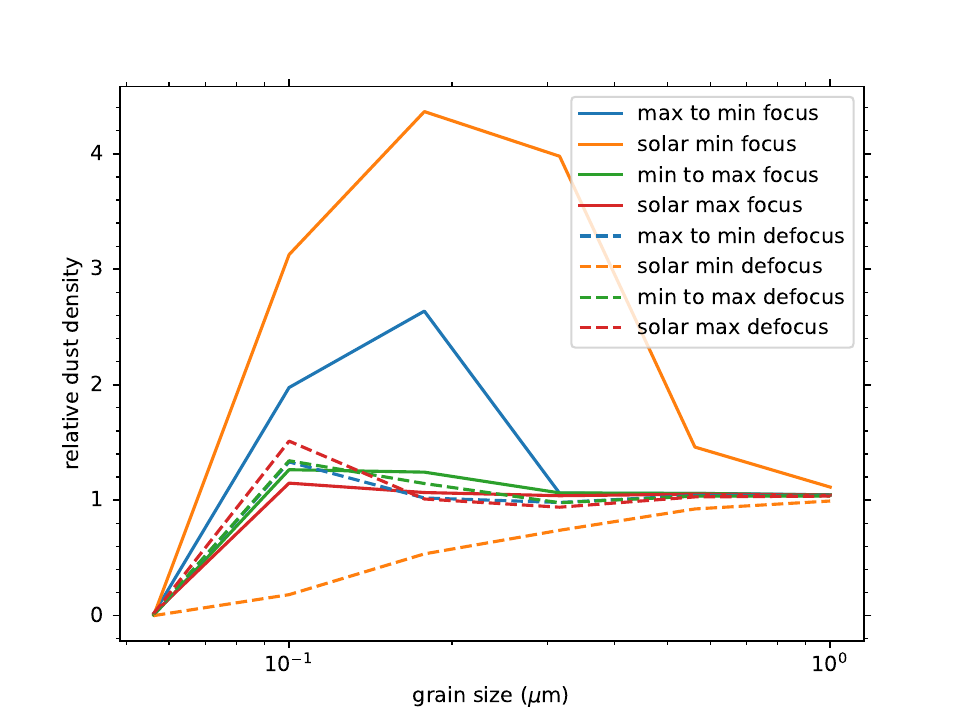}
    \caption{Averaged relative dust density in the region upstream of the Sun.
    Here we show the mean dust density relative to ambient for a volume going
    from the Sun to 20 au upstream ($x$) and along the ecliptic from -10 au to
    +10 au ($y$) and from 5 au above and below the ecliptic ($z$). The
    different lines show results for the focusing polarity (solid) and
    defocusing polarity (dashed) and for the different solar cycle sector
    regions (colors) as a function of grain size. We see that in this region
    and for most grain sizes, there is little deviation from the ambient dust
    density. The exceptions are for solar min conditions (especially for
    focusing polarity), and to a lesser degree the solar max to min transition
    case with focusing polarity.}
    \label{fig:region_avg}
\end{figure}

In Figure \ref{fig:region_avg} we show the relative grain density averaged
over a volume upstream of the Sun. The volume used goes from 0 to $-20$ au in
the upstream direction ($x$), $-10$ to $+10$ au across the ecliptic ($y$) and
$-5$ to $+5$ au perpendicular to the ecliptic plane ($z$). By averaging in
this way we smooth out some of the noise of the density calculations. We see
that for most times for grains with $a \geq 0.1\,\mu$m the grain density is
not strongly disturbed from the ambient density. The exceptions are for solar
min conditions, particularly when the overall polarity is focusing, and for
smaller grains for the transition from solar max to min with overall focusing
polarity. \citet{Strub_etal_2015} found that relative increases in dust flux
of a similar order in the Ulysses data and attributed it to the solar cycle. 
In addition, in all cases, very few grains as small as 0.056 $\mu$m can
penetrate to the inner solar system as can also be seen in Figures
\ref{fig:ddens_small_solmax} and \ref{fig:ddens_small_solmin}. While a flux of
such small particles was observed by Ulysses \citep{Kruger_etal_2015} , it is
well below what would be expected for a typical interstellar grain size
distribution and dust-to-gas ratio.

A direct comparison with previous modeling work is difficult because of
different assumptions, e.g. starting the grains inside of the termination
shock \citep{Sterken_etal_2015} or inside the heliopause
\citep{Czechowski+Mann_2003}.  \citet{Sterken_etal_2015} modeled the regions
sampled by Ulysses including the time dependence of the magnetic field during
transport but not the heliospheric boundaries. They found significant
increases in relative dust density during early mission (their Figure 12,
right panel), similar to our results in Figure \ref{fig:region_avg} for solar
minimum with focusing polarity. They also show substantial depletion of grains
for later periods, which correspond to solar minimum with defocusing polarity,
similar to our results.  Perhaps the most similar calculation to ours was by
\citet{Godenko+Izmodenov_2024}, though in their case the time evolution of the
heliosphere during grain propagation was included. As noted above, this
requires many more trajectories to be followed, since each point is a dust
grain rather than a timestep along a trajectory. They find that 0.15 $\mu$m
grains are deflected by the  heliospheric interface in such a way to enhance
their density at the location of Ulysses trajectory. This differs somewhat
from our results in that for most periods in the solar cycle we find that
grains of roughly that size show little variation from interstellar density.
We do see significant enhancement or depletion of such grains during solar min
conditions. More analysis of the effects of the changing heliospheric
conditions during grain transport would be needed to assess whether this is
the reason for those differences.

\section{Conclusions}
Our goal in this work has been to explore the effects of the sector region on
the transport of interstellar grains through the heliosphere. As others have
found \citep{Czechowski+Mann_2003,Godenko+Izmodenov_2024}, the presence of the
sector region strongly mitigates the solar cycle variations in dust density
that would be expected because of the flipping of the magnetic field polarity
between focusing and de-focusing. It is somewhat surprising that there is such
a strong cutoff in the density, regardless of the point in the solar cycle,
between 0.056 and 0.1 $\mu$m sized grains. Although the sharpness of the
cutoff is not seen in the Ulysses data \citep{Kruger_etal_2015}, there is a
steep dependence of the observed grain flux on grain size with grains as small
as 0.056 $\mu$m down by more than two orders of magnitude from what would be
expected with typical ISM dust-to-gas mass ratio and size distribution
\citep[e.g.][]{Mathis_etal_1977} without filtering.

The morphology of the solar wind magnetic field has a strong influence on the
transport of dust in the heliosphere, especially in the outer heliosphere
where the termination shock and heliopause lead to large departures from the
Parker spiral that predominates in the closer regions of the heliosphere. As
we have shown in this work, and others have shown previously
\citep[e.g.][]{Czechowski+Mann_2003,Godenko+Izmodenov_2024}, the
influence of the sector region is also very important, and largely mitigates
the effects of the compression of the magnetic field in the heliosheath on the
dust transport, in the sense that the heliopause presents much less of a
barrier to entry into the heliosphere. We have shown that during most of the
solar cycle, for grains with $a \gtrsim 0.1\,\mu$m, there is little
compression or dilution of the dust, which is to say that the observed dust
flux is close to that from the undisturbed ISM. However, there are periods
during the cycle, specifically during solar minimum when the sector region is
relatively thin, that grains smaller than $\sim 1\,\mu$m are significantly
focused or de-focused. For those cases a more involved analysis needs to be
done which includes the time dependence of the magnetic field and sector
region which evolve during a grain's transit from the heliopause to near
Earth.

\begin{acknowledgments}
This work was supported by NASA grant 80NSSC22M0164, 18-DRIVE18\_2-0029 as part of the NASA/DRIVE program titled ``Our Heliospheric Shield''. For more information about this center please visit: \url{https://shielddrivecenter.com}.
\end{acknowledgments}


\software{astropy \citep{Astropy_2013}, matplotlib \citep{Hunter:2007}, DVODE \citep{Brown_etal_1989}, BATS-R-US \citep{Toth12}
          }

\bibliography{ISD_in_heliosphere}{}
\bibliographystyle{aasjournal}



\end{document}